\documentclass[twocolumn,nodate,shownopacs,preprintnumbers,nofootinbib,amsmath,amssymb,aps,prl,superscriptaddress]{revtex4-2}

\usepackage{graphicx,wrapfig}

\usepackage{amssymb}

\usepackage{color}
\usepackage{mathrsfs}
\usepackage{amsmath}
\usepackage{subfigure}
\usepackage{afterpage}
\usepackage{multirow}

\usepackage{tikz}
\usepackage{amsfonts}
\usepackage{psfrag}
\usepackage{accents}
\usepackage{enumitem}
\usepackage{hyperref}

\usepackage{caption}
\usepackage{epstopdf}

\DeclareGraphicsExtensions{.pdf,.png,.jpg,.jpeg,.eps}
\def\ig{\includegraphics}

\newcommand{\pb}[1]{\hbox{\lower0.5ex\hbox{${}_{\leftarrow}$}}\kern-1.9ex{#1}}

\def\f{\frac}

\def\t{\tilde}

\def\={\,\hat{=}\,}
\def\Lie{\mathcal{L}}
\def\J{\mathcal{J}}

\def\qo{\mathring{q}}

\def\ko{\mathring{k}}
\def\Do{\mathring{D}}
\def\kappao{\mathring\kappa}
\def\Omegao{\mathring\Omega}

\def\Qko{Q_{{}^{(\ko)}}}
\def\to{\mathring{t}}
\def\varphio{\mathring{\varphi}}
\def\DHS{\mathcal{H}}
\def\qt{\tilde{q}}

\def\QLH{\mathfrak{h}}

\def\kub{\underbar{k}}
\def\tauh{\hat{\tau}}
\def\rh{\hat{r}}

\def\E{\mathcal{E}}
\def\N{\mathcal{N}}
\def\n{\mathfrak{n}}
\def\e{\mathfrak{e}}

\def\rmd{{\rm d}}

\def\pb{\bar{p}}

\def\xiul{\underline{\xi}}

\usepackage{enumerate}

\def\be{\nopagebreak[3]\begin{equation}}
\def\ee{\end{equation}}
\def\ba{\nopagebreak[3]\begin{eqnarray}}
\def\ea{\end{eqnarray}}

\newcommand{\bfig}{\nopagebreak[3]\begin{figure}}
\newcommand{\efig}{\end{figure}}

\newcommand{\bmult}{\nopagebreak[3]\begin{multline}}
\newcommand{\emult}{\end{multline}}

\begin{document}

\title{Thermodynamics of Black Holes, far from Equilibrium}

\author{Abhay Ashtekar}
\affiliation {
Physics Department, Penn State University, University Park, PA 16801}
\author{Daniel E. Paraizo}
\affiliation {
Physics Department, Penn State University, University Park, PA 16801}
\author{Jonathan Shu}
\affiliation{
Physics Department, Penn State University, University Park, PA 16801}

\begin{abstract}

As in thermodynamics, the celebrated first law of black hole mechanics relates infinitesimal changes in the properties of nearby equilibrium states of black holes (without reference to any physical process that causes the transition). Using dynamical horizon segments (DHSs), we extend the first law to encompass black holes that can be arbitrarily far from equilibrium.  It now refers to \emph{finite} changes that occur due to \emph{physical processes}. This extension, together with the generalized second law on DHSs \cite{Ashtekar:2003hk}, naturally lead one to identify entropy of dynamical BHs with the area DHSs.

\end{abstract}

\maketitle

\emph{\bf Introduction}:\,\, Some five decades ago, Bardeen, Carter and Hawking (BCH) \cite{Bardeen:1973gs,LesHouches:1973} discovered the first law for \emph{Killing horizons} (KHs) that governs the relation between two nearby stationary, axisymmetric black holes (BHs) in general relativity. Hawking also showed that, in dynamical situations, the area $A$ of an \emph{event horizon} (EH) cannot decrease if the space-time is asymptotically predictable and satisfies the null energy condition \cite{Hawking:1971vc}. The close similarities with the first and second laws of thermodynamics, and Hawking's subsequent discovery \cite{Hawking:1974sw} that BHs emit quantum radiation at a temperature $(\kappa \hbar)/2\pi$ led to the identification of $A/4G\hbar$ as BH entropy. These discoveries have continued to inspire investigations on the quantum nature of BHs for half a century! 

However, EHs are teleological; they can form and grow in flat regions of space-time (see, e.g.,
\cite{Ashtekar:2004cn,kehle2024extremalblackholeformation,afshordi2024blackholesinside2024}). To determine if a given space-time admits an EH one needs to know the metric all the way to the \emph{infinite future}.
Therefore, their use in \emph{dynamical situations} leads to serious difficulties. At the practical level, one cannot use EHs e.g. in the course of numerical simulations to locate the progenitor BHs, nor the remnant; they can only be introduced as an `afterthought' once the simulation is complete. In the investigation of conceptual issues  --such as cosmic censorship in classical general relativity \cite{Andersson:2007fh}, or the endpoint of the evaporation process of BHs in quantum gravity \cite{Ashtekar_2020}-- one cannot make \emph{a priori} assumptions about the global nature of space-time. Hence, one does not know whether EHs even exist in interesting situations \cite{PhysRevD.36.1065}. What, then, should BH entropy refer to in dynamical situations? 

An answer is provided by quasi-local horizons (QLHs)  
(see, in particular, 
\cite{Hayward:1993wb,Ashtekar:2000hw,Ashtekar:2001is,Ashtekar:2002ag,Ashtekar:2003hk,Ashtekar:2004cn,Booth:2005qc,Gourgoulhon:2005ng,Hayward:2009ji,Jaramillo:2011zw,Ashtekar:2025LivRev}).   
Since they are free of teleology, they are routinely used in numerical simulations of BH mergers.  Their properties have been investigated using geometric analysis (see, e.g.,\cite{Ashtekar:2005ez,Bartnik:2005qj,Andersson:2005gq,Andersson:2007fh,Bengtsson_2011,senovilla2012stabilityoperatormotscore}) as well as numerical methods \cite{Prasad:2020xgr,Prasad:2021dfr,PhysRevD.100.084044,Booth:2021sow,Chen_2022},
and they have also been used to analyze the BH evaporation process (see, e.g.,
\cite{Ashtekar:2005cj,Sawayama:2005mw,Ashtekar_2011a,Ashtekar_2011b,ashtekar2023regularblackholesloop,Agullo_2024}). In this Letter, we will use QLHs to extend the first law to fully dynamical situations. Key new elements are: (i) Introduction of a natural causal vector field $\xi^a$ and the associated notion of energy on horizons of fully dynamical BHs; and, (ii) A conceptual arena that intertwines non-equilibrium and equilibrium states of BHs, making it possible to assign (time dependent) intensive parameters and extensive observables to dynamical BHs. For details, see \cite{aps-detailed}.

\emph{\bf The BCH First Law and its Reformulation}:\,\, In standard treatments, one begins with asymptotically flat space-times $(M, g_{ab})$ that admit a time translation Killing field $\to^a$, a rotational Killing field $\varphio^a$, and a KH $H$ to  which a (constant) linear combination $\ko^a := \to^a +\Omegao\, \varphio^a$ of the two Killing fields serves as the null normal. The freedom in rescaling $\ko^a$ by a constant is fixed by requiring that $\to^a$ be the \emph{unit} time translation at infinity (and the affine parameter of $\varphio^a$ run from $0$ to $2\pi$). 
Einstein's equations imply that a number of identities must hold between geometrical quantities evaluated at $H$ and those evaluated at spatial infinity. Among them is the celebrated BCH first law: $\delta M = (\kappao/8\pi G) \delta A + \Omegao\, \delta J$. Here $M, J$ are taken to be the Arnowitt-Deser-Misner (ADM) quantities, evaluated at infinity, while $A$ is the horizon area, and $\kappao$ the surface gravity of $\ko^a$, both evaluated at the horizon. 

This formulation of the first law has an unsatisfactory feature \cite{Ashtekar:2000hw,Ashtekar:2001is}: the normalization condition on $\to^a$ and the ADM charges $M, J$ refer to space-time geometry near spatial infinity, far from the black hole. One would expect the laws of BH mechanics to refer just to the black hole. Indeed, quantities that enter the first law of thermodynamics refer \emph{only} to the system under consideration. In dynamical situations, one has to consider physical processes that change the mass and angular momentum of the BH. Then the limitation becomes much more severe because the ADM $M$ and $J$ refer to the total system and are \emph{absolutely conserved in physical processes}! $\delta M$ and $\delta J$ in the first law should now refer to infalling energy and angular momentum \emph{across the horizon of each BH} in the given space-time, without reference to the exterior.

Fortunately, one can recast the standard first law using quantities that are specified \emph{just at} $H$ and then extend it to dynamical situations. 
First note that the horizon areal radius $R$ (defined by $A =4\pi R^2$), and the horizon angular momentum $J_H = \f{1}{16\pi G} \oint \epsilon_{abcd}(\nabla^a\varphio^b)\, \rmd S^{cd}$ ($-\f{1}{2}$ the Komar integral of $\varphio^a$ at $H$) are determined by fields \emph{at the horizon}.  Second, a Kerr black hole is fully characterized by values of $(R, J_H)$.  Third, since the surface gravity $\kappao$ scales linearly with the horizon null normal, one can fix the rescaling freedom in $\ko^a$ by requiring that its surface gravity $\kappao$ --also determined by the horizon geometry-- be given by $\kappao = \kappa_{{}_{\rm Kerr}}(R,J_H)$. As noted above, there is no rescaling freedom in $\varphio^a$. Therefore by setting $\Omegao = \Omegao_{{}_{\rm Kerr}}(R,J_H)$, one can recover the restriction of $\to^a$ to $H$ via $\to^a \= \ko^a -\Omegao \varphio^a$. (Throughout $\=$ \emph{will stand for equality at the QLH} under consideration.) Then, the three (Komar) charges \emph{evaluated at the horizon} are given by $\Qko \= \f{\kappao\,A}{4\pi G},\, Q_{(\varphio)} \= -2J_H$\, and\, $Q_{(\to)} \= \f{\kappao\,A}{4\pi G} + 2 \Omegao J_H $. They satisfy the first law 
\be\delta Q_{(\to)} \= \f{\kappao}{8\pi G} \delta A + \Omegao\, \delta J_H\, . \label{BCHlaw}\ee
Since every step now refers just to the horizon, as we show below, the discussion can be extended to dynamical BHs. \medskip

\emph{\bf Dynamical situations}:\,\, For dynamical BHs, there are no KHs, and EHs are unsuitable because their growth is teleological, unrelated to local physical processes (See Fig.1). But we do have \emph{quasi-local horizons} (QLHs) that are free from these limitations.

A QLH $\QLH$ is a 3-manifold that admits a foliation by a 1-parameter family of \emph{marginally trapped surfaces} (MTSs), i.e. closed 2-surfaces on which the expansion $\theta_{(k)}$ of a null normal $k^a$ vanishes identically. In this Letter, we will assume that $\QLH$ is topologically $S^2 \times R$ (see \cite{aps-detailed} for justification). An open portion $\DHS$ of a QLH is said to be a \emph{Dynamical Horizon Segment} (DHS) if:\, (i) It is nowhere null; (ii) the expansion $\theta_{(\kub)}$ of the other null normal $\kub^a$ to MTSs is no where vanishing; and, (iii) a genericity condition holds: $ E:= |\sigma|^2 + R_{ab}k^a k^b$ does not vanish identically on any MTS, where $\sigma_{ab}$ is the shear of $k^a$ and $R_{ab}$ the space-time Ricci tensor. (This condition removes some highly symmetric, degenerate cases \cite{Senovilla_2003,Ashtekar:2025LivRev}.) $E$ represents a `null-energy flux' into the horizon. When this flux vanishes,  the QLH becomes null and is called a \emph{Non Expanding Horizon Segment} (NEHS).  The space-time metric induces on any NEHS a degenerate metric $\qo_{ab}$ of signature $(0,\,+\,+)$, and an intrinsic derivative operator $\Do$. The pair $(\qo_{ab}, \Do)$ is said to constitute the NEHS-geometry. An NEHS is said to be an \emph{Isolated Horizon Segment} (IHS) if it admits a null normal $\ko^a$ that is a symmetry of the horizon geometry, i.e., preserves $(\qo_{ab}, \Do)$. If it does, then $\ko^a$ is unique up to a constant rescaling. On axisymmetric IHSs, one fixes the rescaling freedom by requiring $\kappao\,\=\, \kappa_{\rm kerr}\,(R,\,J)$, as on the Kerr KH \cite{aps-detailed}. Every KH is an IHS. But an IHS is more general; it represents a BH that is itself in equilibrium, allowing for dynamical processes arbitrarily close to it (as, e.g., the Robinson-Trautmann solutions \cite{pc,Podolsky:2009an}). Still, the first law (1) extends to IHSs \cite{aps-detailed,Ashtekar:2001is}. Finally, as Fig. 1 illustrates, QLHs of dynamical BHs can have multiple segments: A DHS followed by an IHS, followed by a DHS, etc. 
\medskip
\bfig \vskip-0.3cm
\hskip-0.5cm
 \ig[width=2.8in,height=1.7in]{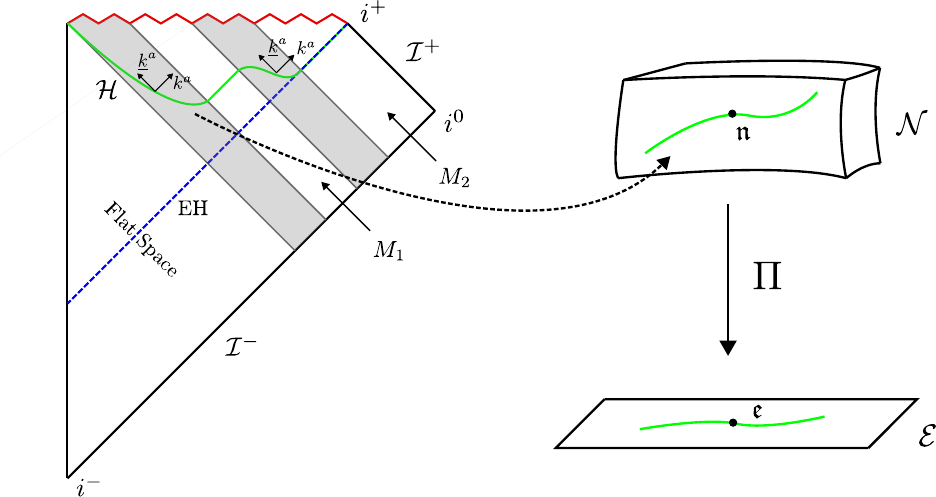}
  \caption{\footnotesize{\emph{A double Vaidya null fluid collapse.} The EH  forms and grows in the flat region of space-time. The QLH (in green) lies entirely in the curved region. It starts out as a (space-like) DHS that grows in area in response to infalling matter, then settles down to a (null) IHS in the space-time region $M_1$, only to become a DHS because of the second infall and finally settles to an IHS in $M_2$, the only segment that coincides with the EH. As the right panel shows, the DHS defines a curve in the space $\N$ of non- equilibrium states that projects down to a curve in the space $\E$ of Kerr equilibrium states under the map $\Pi$.}} 
\vskip-0.5cm
 \label{fig:1} 
\efig

\emph{\bf Overcoming Conceptual Obstacles}:\,\, BHs in general relativity turn out to be both simpler and more complicated than the familiar thermodynamical systems. In standard thermodynamics, a major obstacle in the passage to non-equilibrium situations is that one cannot unambiguously assign intensive parameters such as the temperature or pressure to non-equilibrium states. As we now show, this obstacle can be removed for dynamical BHs in general relativity. 

Recall first that the Kerr equilibrium state $\e$ of a black hole is characterized by two numbers $(R, J_H)$. Thus the space $\E$ of equilibrium states is 2-dimensional. By contrast, the space $\N$ of non-equilibrium states $\n$ is infinite dimensional since each $\n$ is characterized by the plethora of fields on MTSs $S$ of a DHS $\DHS$ (that evolve from one MTS to another). One of them, the 2-metric $\qt_{ab}$ on any $S$, determines its areal radius $R[S]$. For brevity, in this Letter we will restrict ourselves to axisymmetric DHSs. Then, we also have angular momentum: $\J_\DHS[S]\,\=\, -\f{1}{8\pi G}\, \oint_S \kub_b\, \t{q}_a{}^c\,(\nabla_c k^b)\, \rmd^2 V$, which is $-\f{1}{2}$-times the Komar integral, mirroring the relation on the Kerr KH. (As discussed in \cite{aps-detailed}, thanks to results in \cite{Korzynski:2007hu,akkl1,Ashtekar:2021kqj}, the restriction to axisymmetric DHSs is not essential.)

Therefore, there is a natural projection map  {\smash{$\Pi: \N \rightarrow \E$}}, shown in Fig 1, with $\Pi (\n)= \e$ iff  $(R, J_H) =(R[S], \J_\DHS [S])$, that focuses only on values of two observables, ignoring all other information in $\n$. Thanks to the angular momentum-area inequality \cite{Jaramillo:2011pg} for MTSs, this map is well-defined. Via pull-back, we can assign to any $\n$, the intensive parameters $(\kappao,\, \Omegao)$ carried by $\e$.  Thus, thanks to the BH uniqueness theorems, a key obstacle to extending thermodynamics to non-equilibrium situations is overcome: each non-equilibrium state $\n$ has been assigned intensive parameters. Of course, $(\kappao,\, \Omegao)$ are `time-dependent' on $\DHS$, since  $(R[S], \mathcal{J}_{\DHS} [S])$ change from one MTS to another, reflecting the fact that there are physical fluxes traversing $\DHS$. 

However, as remarked earlier, BHs are also more complicated! In thermodynamics, the notion of total energy of a system in a non-equilibrium state is well-defined. In general relativity, on the other hand, the notion of energy is tied to causal vector fields. On KHs, it was natural to use the horizon Killing field $\ko^a$ to define energy/charge $Q_{(\ko)} \,\=\, \f{\kappao\,A}{4\pi G}$. What would be its analog on DHSs? Recall that DHSs carry a preferred null direction field $k^a$ (with $\theta_{(k)} \=0$) which, furthermore, aligns with $\ko^a$ as a DHS reaches equilibrium, becoming an IHS (as in Fig. 1). Hence a natural replacement of $\ko^a$ would be an appropriately scaled vector $\xi^a$ that is parallel to $k^a$. Now, it was shown in \cite{Ashtekar:2003hk} that, thanks to the constraint equations on the DHS, 
there is an invariantly defined charge $Q_{(\xi)} [S]$ associated with any such $\xi^a$. Suppose the infalling flux ends at a MTS $\mathring{S}$ of the QLH, so that the DHS transitions to an IHS at $\mathring{S}$. Then, the DHS and IHS assignments of charges at $\mathring{S}$ would agree \emph{if} $\xi^a$ could be chosen such that the DHS charge $Q_{(\xi)}[\mathring{S}]$ equals the IHS charge $\Qko \= \f{\kappao\,A}{4\pi G}|_{\mathring{S}}$  for any MTS $\mathring{S}$.  
Surprisingly, \emph{this consistency condition selects a unique $\xi^a$ on the DHS.} 

For this $\xi^a$, then, the value of the charge $Q_{(\xi)}[S]$ on any MTS $S$ of $\DHS$, equals the value of the charge $\Qko$ evaluated at the equilibrium state $\e$ that the projection map $\Pi$ assigns to $S$. By the very definition of $\Pi$,\, the agreement also holds between $\mathcal{J}_{\DHS} [S]$ and $J_H|_{\e}$. Thus, $\Pi$ assigns to each DHS $\DHS$, a trajectory $\e(R)$ in the space $\E$ of equilibrium states (see Fig. 1), such that values of  charges $(Q_{(\xi)}[S],\, \mathcal{J}_{\DHS}[S])$ on $\DHS$ agree with the values of $(\Qko,\, J_H)|_{\e}$ all along $\e(R)$! This surprising synergy between values of observables defined \emph{intrinsically} on $\DHS$ and those on $\E$ enable one to probe dynamics using tools from equilibrium thermodynamics of BHs.
\medskip

\emph{\bf 1st and 2nd laws on DHSs}:\,\, For definiteness, let us first focus on space-like DHSs. Then, $\tauh^a$ the unit normal to a DHS $\DHS$ is time-like; $\rh^a$ the unit normal within $\DHS$ to MTSs $S$ is space-like; and $k^a \=\f{1}{\sqrt{2}}(\tauh^a + \rh^a)$ and $\kub^a \= \f{1}{\sqrt{2}}\,(\tauh^a - \rh^a)$ are the two null normals to $S$. As in the BCH analysis, we will use the 4 constraint equations, \emph{but now on the DHS} $\DHS$ rather than on a partial Cauchy surface extending from the horizon to spatial infinity. Specifically, we will use combination $ (G_{ab} - 8\pi G\, T_{ab})\, \xi^a \tauh^b \=0$\, of constraints, with $\xi^a \= \sqrt{2}\, |DR| f(R)\, k^a$, choosing functions $f(R)$ appropriately. (For motivation, see \cite{Ashtekar:2003hk}.)

On the DHS $\DHS$, $\mathcal{J}_{\DHS}[S]$ is a function of $R$, whence $\kappao$ and $\Omegao$ are also functions of $R$, reflecting their dynamical nature. The unique choice of $\xi^a$ for which  (the $\xi$-energy charge) $Q_{(\xi)}[S]\, \= Q_{(\ko)} \equiv \f{\kappao\, A}{4\pi G}$, is given by setting $f(R) \= \f{\rmd}{\rmd R}\, (\f{\kappao\, A}{2\pi G})$. Constraint equations on $\DHS$ provides a balance law for this $Q_{(\xi)}$. Setting {\smash{$\Delta\, [Q_{(\xi)}]\, \= Q_{(\xi)}[S_2] - Q_{(\xi)}[S_1]$,} one has  
\be\Delta\, [Q_{(\xi)}]\, \= \,\mathfrak{F}_{(\xi)} [\DHS_{1}^{2}]\, \equiv\, \int_{\DHS_1^2} \big(\mathfrak{f}_{(\xi)}^{\,\,\rm matt} + \mathfrak{f}_{(\xi)}^{\,\,\rm gws}\big) \rmd^3 V. \label{balance1}\ee
\noindent Here $\DHS_1^2$ is the portion of the DHS bounded by MTSs $S_1$ and $S_2$, and $\mathfrak{F}_{(\xi)}$ the flux of $\xi$-energy, with flux densities for matter fields and gravitational waves given by:
\vskip0.2cm
\centerline{$\mathfrak{f}_{(\xi)}^{\,\,\rm matt}  \= T_{ab}\, \xi^a \tauh^b$\,\, and}\vskip0.1cm
\centerline{$\mathfrak{f}_{(\xi)}^{\,\,\rm gws}  \= \f{1}{8\pi G}\, |DR|\,f(R)\, (|\sigma|^2 + 2 |\zeta|^2)$ ,}
\vskip0.2cm
\noindent with $\sigma_{ab}$, the shear of the null normal $k^a$, and $\zeta^a \=\, \qt^{ab}\, \rh^c \nabla_c k_b$. The appearance of $|\zeta|^2$ seems surprising at first because it is absent in perturbative expressions of the energy carried by gravitational waves across KHs. It arises because $\DHS$ is space-like rather than null: While there are only 2 true \emph{phase space} degrees of freedom on null surfaces in general relativity, on a space-like surface there are 4, now captured in the fields $(\sigma_{ab}, \, \zeta^a)$ tangential to MTSs, each with 2 free components \cite{Ashtekar:2025LivRev}. They all contribute to the energy flux, as is usual in field theories.

To obtain the dynamical version of the first law (\ref{BCHlaw}), let us set $t^a := \xi^a - \Omegao \varphi^a$ on $\DHS$ (mimicking $\to^a = \ko^a - \Omegao \varphio^a$ on the Kerr KH). Then, the balance law (\ref{balance1}) is recast as:
\vskip-0.1cm
\be \Delta\, [Q_{(\xi)}] \equiv \Delta\, [\frac{\kappao\,A}{4\pi G}]\, \=\, \mathfrak{F}_{(t)} [\DHS_{1}^{2}] + \mathfrak{F}_{(\Omegao \varphi )} [\DHS_{1}^{2}]. \label{balance2}\ee
\noindent Matter contributions to the fluxes on the right side are obtained by replacing $\xi^a$ in $\mathfrak{f}_{(\xi)}^{\,\,\rm matt}$ by $t^a$ and $\Omegao \varphi^a$ respectively. The gravitational contributions are given by \cite{Ashtekar:2003hk}:\vskip0.2cm
\centerline{ $\mathfrak{f}_{(\Omegao \varphi )}^{\,\,\rm gws}   \=
 \f{1}{8\pi G} (K^{ab} - K q^{ab})\, (\Lie_{(\Omegao\, \varphi)}\, q_{ab})$,}  
\vskip0.2cm
\noindent leading to the equality $\mathfrak{F}_{(\Omegao \varphi )} \= \Delta [Q_{(\Omegao\varphi)}]$.  Setting
\vskip0.2cm
\centerline{ $\mathfrak{f}_{(t)}^{\,\,\rm gws}\, :=\, \mathfrak{f}_{(\xi)}^{\,\,\rm gws} - \mathfrak{f}_{(\Omegao \varphi )}^{\,\,\rm gws}$,}
\vskip0.2cm
\noindent and $Q_{(t)} := Q_{(\xi)} - Q_{(\Omegao\varphi)}$, we have $\mathfrak{F}_{(t)}\= \Delta [Q_{(t)}]$.} Since $Q_{(\Omegao\varphi)} \= -2\J_\DHS$, the balance law (\ref{balance2}) is equivalent to:
\be  \Delta\, [Q_{(t)}]\, \=\, \Delta\, [\f{\kappao A}{4\pi G}]\, +\, 2 \Delta\, [\Omegao \mathcal{J}_{\DHS}]. \label{1law}\ee
\noindent This is our dynamical generalization of the first law (\ref{BCHlaw}) to DHSs. There are some notable differences between (\ref{1law}) and (\ref{BCHlaw}). First, $\Delta$ refers to \emph{finite} changes. Second, $\Delta Q_{(t)}$ and $\Delta (\Omegao \mathcal{J}_{\DHS})$ equal the \emph{fluxes} of $t$-energy and $(\Omegao\varphi)$-\,angular momentum across $\DHS_1^2$ due to \emph{physical processes} in space-time; this is an \emph{active} version of the first law. Third, $\kappao$ and $\Omegao$ are now `time-dependent' --they change from one MTS to another-- and therefore do not come out of $\Delta$. Fourth, the right side differs from that of the KH first law (\ref{BCHlaw}) by a factor of 2. 

The third and fourth differences are intertwined. This becomes transparent in the infinitesimal form of the integral version, obtained by letting $S_2$ approach $S_1$ so that they are infinitesimally separated. Then Eq. (\ref{1law}) reduces to $\delta Q_{(t)}\, \=\, \delta [\f{\kappao A}{4\pi G}]\, +\, 2 \delta [\Omegao \mathcal{J}_{\DHS}]$, where $\delta$ denotes infinitesimal changes at $S_1$, along $\DHS$. Since every DHS defines a trajectory\, $\e(R)$\, on the space $\mathcal{E}$ of equilibrium states, $S_1$ corresponds to a point $\e_1$ in $\E$. Thanks to the properties of  the projection $\Pi$, the infinitesimal changes $\delta Q_{(t)},\, \delta [\f{\kappao A}{4\pi G}]$ and $\delta [\Omegao \mathcal{J}_{\DHS}]$ on $\DHS$ equal infinitesimal changes $\delta Q_{(\to)},\, \delta [\f{\kappao A}{4\pi G}]$ and $\delta [\Omegao J_H]$  at $\e_1$, along $\e(R)$. Now,  the Smarr relations on the KH imply an identity on $\E$: $\f{A\, \delta \kappao}{4\pi G} + 2J_H \delta\Omegao \= -\big(\f{\kappao\, \delta A}{8\pi G} + \Omegao \delta J\big)$. By pulling it back to $\DHS$ via $\Pi$, the infinitesimal version of the DHS first law can be rewritten as
%
\be \delta Q_{(t)} \= \f{\kappao}{8\pi G}\,\delta A\,+\, \Omegao\, \delta \mathcal{J}_{\DHS}\, , \ee 
%
\noindent which has exactly the same form as the first law (\ref{BCHlaw}) on KHs. (But on the DHS the infinitesimal changes are caused by physical fluxes into $\DHS$ (near $S_1$); it is not a passive change from one equilibrium state to a nearby one.) The finite version (\ref{1law}) of the first law cannot be rewritten in this way:  since $(\kappao, \Omegao)$ are dynamical on $\DHS$, they vary along the trajectory\, $\e(R)$\, in $\E$, whence the identity we used does not hold for finite variations.

This concludes our discussion of the generalization of the first law to dynamical BHs in general relativity. For completeness, we will recall from \cite{Ashtekar:2003hk} the second law on DHSs, to bring out its synergy with the first law. The second law also results from the combination $(G_{ab} - 8\pi G\, T_{ab}){\xiul}^a \tauh^b \=0$ where now the multiple of $k^a$ is now given by  ${\xiul}^a := \sqrt{2} |DR| (4\pi R)\, k^a$. This constraint implies: 
\be \f{\Delta A}{4G} \= \int_{\DHS_1^2}\, \big[T_{ab}\xiul^a \tauh^b + \f{R |DR|}{2G}(|\sigma|^2 + 2|\zeta|^2)\big]\, \rmd^3 V. \ee
\noindent If matter satisfies the dominant energy condition \emph{at} $\DHS$, the right side is positive definite, whence the area increases. But in contrast to Hawking's result for EHs \cite{Hawking:1971vc}, the statement of the second law on DHSs is quantitative: \emph{It is directly related to the influx of energy across the segment $\DHS_1^2$ of the DHS}. As Fig. 1 shows, the area of EHs can increase across a segment even when the segment lies in a flat region of space-time where nothing is happening!
\vskip0.1cm

\emph{\bf Discussion}:\, Space-times representing dynamical BHs do not admit KHs. They do admit EHs, but  the growth in the area of EHs is unrelated to local physical processes. Therefore, the EH area \emph{cannot} be a viable measure of the physical entropy in non-equilibrium situations. DHSs share some attractive properties with EHs --e.g., their area grows in classical GR, and they can merge but cannot bifurcate \cite{Ashtekar:2005ez}. Furthermore, DHSs are defined quasi-locally and are free from teleology. Therefore, they have been widely used to represent black holes far from equilibrium both in classical and quantum gravity. In this Letter we showed that they are also well-suited for extending BH thermodynamics to non-equilibrium situations. In particular, the expressions of the first and second law, and the fact that the time dependence of the area of MTSs of a DHS is governed by \emph{local} physical processes, naturally lead one to identify \emph{the area of MTSs of a DHS with BH entropy} in non-equilibrium situations. Finally, note that our analysis used structures that are directly available in the physical space-time. For example, one does not need bifurcate horizons that are often used, but do not exist in physical space-times of dynamical BHs beyond perturbation theory. 

We will conclude with a few remarks. For further elaboration, see \cite{aps-detailed}.

1. New elements of our analysis can be summarized as follows. While several balance laws have been available in the literature, none was a direct analog of the BCH first law (1). In particular, only one of them has terms of dimensions of energy as in (1). But its left side features the Hawking-mass, which does not in general agree with the charge $Q_{(\to)}$ on IHSs when the BH reaches equilibrium. By contrast, the left side of the new balance law (4) features $Q_{(t)}$, which is free of this drawback. Also, our vector field $\xi^a$ and the associated charge are new constructs that play an essential role in relating DHSs to the recent perturbative results. Finally, the surprising extent of the interplay between equilibrium and non-equilibrium thermodynamics has not been discussed before.

2. For simplicity we focused on space-like DHSs --the most common occurrence in the classical theory \cite{Ashtekar:2025LivRev}. The analysis goes through \cite{Ashtekar:2003hk,aps-detailed} also for time-like DHSs that, e.g., represent evaporating BHs during their long semi-classical phase  \cite{Sawayama:2005mw,Ashtekar_2011b,Ashtekar_2020,ashtekar2023regularblackholesloop,Agullo_2024,Varadarajan:2024clw}. But some of the signs in various expressions flip --e.g. because of the negative energy flux into $\DHS$ during evaporation-- causing the area to decrease.

3. We used DHSs to represent BHs that are not in equilibrium. Generically, in each of its dynamical phases, the BH can admit a number of `interweaving' DHSs $\DHS$ \cite{Ashtekar:2005ez}. A key point is that each $\DHS$ provides us with a detailed, self-consistent  description of the evolution of the BH and all our thermodynamical considerations also hold for each $\DHS$. When the BH reaches equilibrium, represented by an IHS, all these DHSs asymptote to that IHS and their thermodynamic parameters agree with those of the IHS \cite{Ashtekar:2013qta}. Thus, there is consistency between the dynamical and equilibrium phases.

4. Recently it was shown that, already when first order, dynamical perturbations are included around stationary BHs in classical GR, entropy is given, \emph{not} with the area of the perturbed EH, but by areas of a family of MTSs that lie \emph{inside the} EH \cite{PhysRevD.108.044069,hollands2024entropydynamicalblackholes,visser2025dynamicalentropychargedblack}. These results can be obtained using perturbed IHSs \cite{akkl1,Ashtekar:2021kqj}. The world-tube of these MTSs can also be thought of as a `slowly evolving DHS'  \cite{Booth:2003ji}. But conceptually the perturbative and the DHS frameworks are rather different. Similarities and differences are discussed in \cite{aps-detailed}. However, so far our analysis has been focused on QLHs in general relativity. QLHs do exist in other gravitational theories and some of their key properties follow simply from the geometric Gauss-Codazzi equations \cite{Ashtekar:2025LivRev}. It would be interesting to investigate whether ideas underlying our analysis can be extended to other metric theories of gravity. 

{\bf Acknowledgments}: AA acknowledges stimulating discussions on quasi-local horizons with a large number of colleagues over the years, especially Badri Krishnan. Presentation of the material benefited from discussions at the Simons workshop \emph{50 years of black hole information paradox} at Stony Brook, the conference \emph{Cosmological Olentzero} at Bilbao, and an International Loop Quantum Gravity Seminar, particularly the subsequent correspondence with Simone Speziale. This work was supported in part by the Atherton and Eberly funds of Penn State.  D.E.P. acknowledges support via Penn State's Bunton-Waller Award and the Walker Fellowship from its Applied Research Laboratory.

\section*{References}
\bibliography{apslett}

@article{Varadarajan:2024clw,
    author = "Varadarajan, Madhavan",
    title = "{Spherical collapse and black hole evaporation}",
    eprint = "2406.09176",
    archivePrefix = "arXiv",
    primaryClass = "gr-qc",
    doi = "10.1103/PhysRevD.111.026005",
    journal = "Phys. Rev. D",
    volume = "111",
    number = "2",
    pages = "026005",
    year = "2025"
}

@misc{aps-detailed,
      title={Thermodynamics of dynamical black holes beyond perturbation theory}, 
      author={Abhay Ashtekar and Daniel E. Paraizo and Jonathan Shu},
      year={2025},
      eprint={2604.00170},
      archivePrefix={arXiv},
      primaryClass={gr-qc},
      url={https://arxiv.org/abs/2604.00170}, 
}

@article{PhysRevD.108.044069,
  title = {Second law from the Noether current on null hypersurfaces},
  author = {Rignon-Bret, Antoine},
  eprint={2303.07262},
      archivePrefix={arXiv},
      primaryClass={gr-qc},
      url={https://arxiv.org/abs/2303.07262}, 
      journal = {Phys. Rev. D},
  volume = {108},
  issue = {4},
  pages = {044069},
  numpages = {28},
  year = {2023},
  month = {Aug},
  publisher = {American Physical Society},
  doi = {10.1103/PhysRevD.108.044069},
  }

@article{Agullo_2024,
   title={Entangled pairs in evaporating black holes without event horizons},
   volume={110},
   ISSN={2470-0029},
   url={http://dx.doi.org/10.1103/PhysRevD.110.085002},
   DOI={10.1103/physrevd.110.085002},
   number={8},
   journal={Physical Review D},
   publisher={American Physical Society (APS)},
   author={Agullo, Ivan and Calizaya Cabrera, Paula and Elizaga Navascu\'es, Beatriz},
   year={2024},
   month=oct }

@article{Ashtekar_2011a,
   title={Surprises in the Evaporation of 2D Black Holes},
   volume={106},
   ISSN={1079-7114},
   url={http://dx.doi.org/10.1103/PhysRevLett.106.161303},
   DOI={10.1103/physrevlett.106.161303},
   number={16},
   journal={Physical Review Letters},
   publisher={American Physical Society (APS)},
   author={Ashtekar, Abhay and Pretorius, Frans and Ramazanoglu, Fethi M.},
   year={2011},
   month=apr }

@article{Ashtekar_2011b,
   title={Evaporation of two-dimensional black holes},
   volume={83},
   ISSN={1550-2368},
   url={http://dx.doi.org/10.1103/PhysRevD.83.044040},
   DOI={10.1103/physrevd.83.044040},
   number={4},
   journal={Physical Review D},
   publisher={American Physical Society (APS)},
   author={Ashtekar, Abhay and Pretorius, Frans and Ramazanoglu, Fethi M.},
   year={2011},
   month=feb }

@misc{hollands2024entropydynamicalblackholes,
      title={The Entropy of Dynamical Black Holes}, 
      author={Stefan Hollands and Robert M. Wald and Victor G. Zhang},
      year={2024},
      eprint={2402.00818},
      archivePrefix={arXiv},
      primaryClass={hep-th},
      url={https://arxiv.org/abs/2402.00818}, }

@misc{visser2025dynamicalentropychargedblack,
      title={Dynamical entropy of charged black objects}, 
      author={Manus R. Visser and Zihan Yan},
      year={2025},
      eprint={2510.20747},
      archivePrefix={arXiv},
      primaryClass={hep-th},
      url={https://arxiv.org/abs/2510.20747}, 
}

@Article{Ashtekar:2025LivRev,
     author    = "Ashtekar, Abhay and Krishnan, Badri",
     title     = "{Quasi-Local Horizons: Recent Developments}",
     journal   = "Living Rev. Rel.",
     volume    = "28",
     year      = "2025",
     pages     = "8",
     DOI       ={10.1007/s41114-025-00061-4},
     eprint    = "2502.11825",
     archivePrefix = "arXiv",
     SLACcitation  = "%%CITATION = GR-QC/0407042;%%"
}

@article{PhysRevD.100.084044,
  title = {Self-intersecting marginally outer trapped surfaces},
  author = {Pook-Kolb, Daniel and Birnholtz, Ofek and Krishnan, Badri and Schnetter, Erik},
  journal = {Phys. Rev. D},
  volume = {100},
  issue = {8},
  pages = {084044},
  numpages = {14},
  year = {2019},
  month = {Oct},
  publisher = {American Physical Society},
  doi = {10.1103/PhysRevD.100.084044},
  url = {https://link.aps.org/doi/10.1103/PhysRevD.100.084044}
}

@Book{LesHouches:1973,
      author         = "DeWitt, Cecile and DeWitt, Bryce S.", 
      title          = "Black Holes: 23rd session of summer school of Les Houches", 
      year           = 1973,
      publisher      = "Gordon Breach (New York)"
}

@article{Andersson:2007fh,
      author         = "Andersson, Lars and Mars, Marc and Simon, Walter",
      title          = "{Stability of marginally outer trapped surfaces and
                        existence of marginally outer trapped tubes}",
      journal        = "Adv.Theor.Math.Phys.",
      volume         = "12",
      year           = "2008",
      eprint         = "0704.2889",
      archivePrefix  = "arXiv",
      primaryClass   = "gr-qc",
      SLACcitation   = "%%CITATION = ARXIV:0704.2889;%%",
}

@Article{Ashtekar:2001is,
     author    = "Ashtekar, Abhay and Beetle, Christopher and Lewandowski,
                  Jerzy",
     title     = "{Mechanics of Rotating Isolated Horizons}",
     journal   = "Phys. Rev.",
     volume    = "D64",
     year      = "2001",
     pages     = "044016",
     eprint    = "gr-qc/0103026",
     archivePrefix = "arXiv",
     doi       = "10.1103/PhysRevD.64.044016",
     SLACcitation  = "%%CITATION = GR-QC/0103026;%%"
}

@Article{Ashtekar:2000hw,
     author    = "Ashtekar, Abhay and Fairhurst, Stephen and Krishnan, Badri
                  ",
     title     = "{Isolated horizons: Hamiltonian evolution and the first
                  law}",
     journal   = "Phys. Rev.",
     volume    = "D62",
     year      = "2000",
     pages     = "104025",
     eprint    = "gr-qc/0005083",
     archivePrefix = "arXiv",
     doi       = "10.1103/PhysRevD.62.104025",
     SLACcitation  = "%%CITATION = GR-QC/0005083;%%"
}

@Article{Ashtekar:2004cn,
     author    = "Ashtekar, Abhay and Krishnan, Badri",
     title     = "{Isolated and dynamical horizons and their applications}",
     journal   = "Living Rev. Rel.",
     volume    = "7",
     year      = "2004",
     pages     = "10",
     eprint    = "gr-qc/0407042",
     archivePrefix = "arXiv",
     SLACcitation  = "%%CITATION = GR-QC/0407042;%%"
}

@Article{Ashtekar:2002ag,
     author    = "Ashtekar, Abhay and Krishnan, Badri",
     title     = "{Dynamical horizons: Energy, angular momentum, fluxes and
                  balance laws}",
     journal   = "Phys. Rev. Lett.",
     volume    = "89",
     year      = "2002",
     pages     = "261101",
     eprint    = "gr-qc/0207080",
     archivePrefix = "arXiv",
     doi       = "10.1103/PhysRevLett.89.261101",
     SLACcitation  = "%%CITATION = GR-QC/0207080;%%"
}

@Article{Ashtekar:2003hk,
     author    = "Ashtekar, Abhay and Krishnan, Badri",
     title     = "{Dynamical horizons and their properties}",
     journal   = "Phys. Rev.",
     volume    = "D68",
     year      = "2003",
     pages     = "104030",
     eprint    = "gr-qc/0308033",
     archivePrefix = "arXiv",
     doi       = "10.1103/PhysRevD.68.104030",
     SLACcitation  = "%%CITATION = GR-QC/0308033;%%"
}

@article{PhysRevD.36.1065,
  title = {Origin of {H}awking radiation},
  author = {Hajicek, Petr},
  journal = {Phys. Rev. D},
  volume = {36},
  issue = {4},
  pages = {1065--1079},
  numpages = {0},
  year = {1987},
  month = {Aug},
  publisher = {American Physical Society},
  doi = {10.1103/PhysRevD.36.1065},
  url = {https://link.aps.org/doi/10.1103/PhysRevD.36.1065}
}

@Article{Bardeen:1973gs,
     author    = "Bardeen, James M. and Carter, Brandon and Hawking, Stephen W.",
     title     = "{The Four laws of black hole mechanics}",
     journal   = "Commun. Math. Phys.",
     volume    = "31",
     year      = "1973",
     pages     = "161-170",
     doi       = "10.1007/BF01645742",
     SLACcitation  = "%%CITATION = CMPHA,31,161;%%"
}

@Article{Booth:2003ji,
     author    = "Booth, Ivan and Fairhurst, Stephen",
     title     = "{The first law for slowly evolving horizons}",
     journal   = "Phys. Rev. Lett.",
     volume    = "92",
     year      = "2004",
     pages     = "011102",
     eprint    = "gr-qc/0307087",
     archivePrefix = "arXiv",
     doi       = "10.1103/PhysRevLett.92.011102",
     SLACcitation  = "%%CITATION = GR-QC/0307087;%%"
}

@Article{Hawking:1974sw,
     author    = "Hawking, Stephen W.",
     title     = "{Particle Creation by Black Holes}",
     journal   = "Commun. Math. Phys.",
     volume    = "43",
     year      = "1975",
     pages     = "199-220",
     doi       = "10.1007/BF02345020",
     SLACcitation  = "%%CITATION = CMPHA,43,199;%%"
}

@article{Hayward:1993wb,
      author         = "Hayward, Sean A.",
      eprint          = "gr-qc/9303006", 
      archivePrefix =  "arXiv",
      title          = "{General laws of black hole dynamics}",
      journal        = "Phys. Rev.",
      volume         = "D49",
      pages          = "6467-6474",
      doi            = "10.1103/PhysRevD.49.6467",
      year           = "1994",
      SLACcitation   = "%%CITATION = PHRVA,D49,6467;%%",
}

@article{Senovilla_2003,
   title={On the existence of horizons in spacetimes with vanishing curvature invariants},
   volume={2003},
   ISSN={1029-8479},
   url={http://dx.doi.org/10.1088/1126-6708/2003/11/046},
   DOI={10.1088/1126-6708/2003/11/046},
   number={11},
   journal={Journal of High Energy Physics},
   publisher={Springer Science and Business Media LLC},
   author={Senovilla, José M. M},
   year={2003},
   month=nov, pages={046?046} 
}

@Article{Ashtekar:2005ez,
     author    = "Ashtekar, Abhay and Galloway, Gregory J.",
     title     = "{Some uniqueness results for dynamical horizons}",
     journal   = "Adv. Theor. Math. Phys.",
     volume    = "9",
     year      = "2005",
     pages     = "1-30",
     eprint    = "gr-qc/0503109",
     archivePrefix = "arXiv",
     SLACcitation  = "%%CITATION = GR-QC/0503109;%%"
}

@Article{Ashtekar:2005cj,
     author    = "Ashtekar, Abhay and Bojowald, Martin",
     title     = "{Black hole evaporation: A paradigm}",
     journal   = "Class. Quant. Grav.",
     volume    = "22",
     year      = "2005",
     pages     = "3349-3362",
     eprint    = "gr-qc/0504029",
     archivePrefix = "arXiv",
     doi       = "10.1088/0264-9381/22/16/014",
     SLACcitation  = "%%CITATION = GR-QC/0504029;%%"
}

@Article{Bartnik:2005qj,
     author    = "Bartnik, Robert and Isenberg, James",
     title     = "{Spherically symmetric dynamical horizons}",
     journal   = "Class. Quant. Grav.",
     volume    = "23",
     year      = "2006",
     pages     = "2559-2570",
     eprint    = "gr-qc/0512091",
     archivePrefix = "arXiv",
     doi       = "10.1088/0264-9381/23/7/020",
     SLACcitation  = "%%CITATION = GR-QC/0512091;%%"
}

@Article{Booth:2005qc,
     author    = "Booth, Ivan",
     title     = "{Black hole boundaries}",
     journal   = "Can. J. Phys.",
     volume    = "83",
     year      = "2005",
     pages     = "1073-1099",
     eprint    = "gr-qc/0508107",
     archivePrefix = "arXiv",
     doi       = "10.1139/p05-063",
     SLACcitation  = "%%CITATION = GR-QC/0508107;%%"
}

@Article{Gourgoulhon:2005ng,
     author    = "Gourgoulhon, Eric and Jaramillo, Jose Luis",
     title     = "{A 3+1 perspective on null hypersurfaces and isolated
                  horizons}",
     journal   = "Phys. Rept.",
     volume    = "423",
     year      = "2006",
     pages     = "159-294",
     eprint    = "gr-qc/0503113",
     archivePrefix = "arXiv",
     doi       = "10.1016/j.physrep.2005.10.005",
     SLACcitation  = "%%CITATION = GR-QC/0503113;%%"
}

@Article{Hayward:2009ji,
     author    = "Hayward, Sean A.",
     title     = "{Involute, minimal, outer and increasingly trapped
                  spheres}",
     journal   = "Phys. Rev.",
     volume    = "D81",
     year      = "2010",
     pages     = "024037",
     eprint    = "0905.3950",
     archivePrefix = "arXiv",
     primaryClass  =  "gr-qc",
     doi       = "10.1103/PhysRevD.81.024037",
     SLACcitation  = "%%CITATION = 0905.3950;%%"
}

@article{Jaramillo:2011pg,
      author         = "Jaramillo, Jose Luis and Reiris, Martin and Dain, Sergio",
      title          = "{Black hole Area-Angular momentum inequality in
                        non-vacuum spacetimes}",
      journal        = "Phys.Rev.",
      volume         = "D84",
      pages          = "121503",
      doi            = "10.1103/PhysRevD.84.121503",
      year           = "2011",
      eprint         = "1106.3743",
      archivePrefix  = "arXiv",
      primaryClass   = "gr-qc",
      SLACcitation   = "%%CITATION = ARXIV:1106.3743;%%",
}

@Article{Korzynski:2007hu,
     author    = "Korzynski, Mikolaj",
     title     = "{Quasi--local angular momentum of non--symmetric isolated
                  and dynamical horizons from the conformal decomposition of
                  the metric}",
     journal   = "Class. Quant. Grav.",
     volume    = "24",
     year      = "2007",
     pages     = "5935-5944",
     eprint    = "0707.2824",
     archivePrefix = "arXiv",
     primaryClass  =  "gr-qc",
     doi       = "10.1088/0264-9381/24/23/015",
     SLACcitation  = "%%CITATION = 0707.2824;%%"
}

@Article{Sawayama:2005mw,
     author    = "Sawayama, Shintaro",
     title     = "{Dynamical horizon of evaporating black hole in Vaidya
                  spacetime. or:  Evaporating dynamical horizon with Hawking
                  effect in Vaidya spacetime}",
     journal   = "Phys. Rev.",
     volume    = "D73",
     year      = "2006",
     pages     = "064024",
     eprint    = "gr-qc/0509048",
     archivePrefix = "arXiv",
     doi       = "10.1103/PhysRevD.73.064024",
     SLACcitation  = "%%CITATION = GR-QC/0509048;%%"
}

@article{Ashtekar:2013qta,
      author         = "Ashtekar, Abhay and Campiglia, Miguel and Shah, Samir",
      title          = "{Dynamical Black Holes: Approach to the Final State}",
      journal        = "Phys. Rev.",
      volume         = "D88",
      year           = "2013",
      number         = "6",
      pages          = "064045",
      doi            = "10.1103/PhysRevD.88.064045",
      eprint         = "1306.5697",
      archivePrefix  = "arXiv",
      primaryClass   = "gr-qc",
      reportNumber   = "IGC-13-06-2",
      SLACcitation   = "%%CITATION = ARXIV:1306.5697;%%"
}

@article{Jaramillo:2011zw,
      author         = "Jaramillo, Jose Luis",
      title          = "{An introduction to local Black Hole horizons in the 3+1
                        approach to General Relativity}",
      journal        = "Int. J. Mod. Phys.",
      volume         = "D20",
      year           = "2011",
      pages          = "2169",
      doi            = "10.1142/S0218271811020366",
      eprint         = "1108.2408",
      archivePrefix  = "arXiv",
      primaryClass   = "gr-qc",
      SLACcitation   = "%%CITATION = ARXIV:1108.2408;%%"
}

@article{Prasad:2020xgr,
    author = "Prasad, Vaishak and Gupta, Anshu and Bose, Sukanta and Krishnan, Badri and Schnetter, Erik",
    title = "{News from horizons in binary black hole mergers}",
    eprint = "2003.06215",
    archivePrefix = "arXiv",
    primaryClass = "gr-qc",
    reportNumber = "LIGO Preprint number LIGO-P2000098",
    doi = "10.1103/PhysRevLett.125.121101",
    journal = "Phys. Rev. Lett.",
    volume = "125",
    number = "12",
    pages = "121101",
    year = "2020"
}

@article{Prasad:2021dfr,
    author = "Prasad, Vaishak and Gupta, Anshu and Bose, Sukanta and Krishnan, Badri",
    title = "{Tidal deformation of dynamical horizons in binary black hole mergers}",
    eprint = "2106.02595",
    archivePrefix = "arXiv",
    primaryClass = "gr-qc",
    reportNumber = "This manuscript has been assigned the LIGO Preprint number
  LIGO-P2100109",
    doi = "10.1103/PhysRevD.105.044019",
    journal = "Phys. Rev. D",
    volume = "105",
    number = "4",
    pages = "044019",
    year = "2022"
}

@article{akkl1,
	archiveprefix = {arXiv},
	author = {Abhay Ashtekar and Neev Khera and Maciej Kolanowski and Jerzy Lewandowski},
	doi = {10.1007/JHEP01(2022)028},
	eprint = {2111.07873},
	journal = {JHEP},
	number = {1},
	primaryclass = {gr-qc},
	title = {Non-Expanding horizons: Multipoles and the Symmetry Group},
	volume = {2022},
	year = {2022}
	}

@article{Ashtekar:2021kqj,
   title={Charges and fluxes on (perturbed) non-expanding horizons},
   volume={2022},
   ISSN={1029-8479},
   url={http://dx.doi.org/10.1007/JHEP02(2022)066},
   DOI={10.1007/jhep02(2022)066},
   number={2},
   journal={Journal of High Energy Physics},
   publisher={Springer Science and Business Media LLC},
   author={Ashtekar, Abhay and Khera, Neev and Kolanowski, Maciej and Lewandowski, Jerzy},
   year={2022},
   month=feb }

@article{Ashtekar_2020,
   title={Black Hole Evaporation: A Perspective from Loop Quantum Gravity},
   eprint = "2001.08833",
    archivePrefix = "arXiv",
    primaryClass = "gr-qc",
    volume={6},
   ISSN={2218-1997},
   url={http://dx.doi.org/10.3390/universe6020021},
   DOI={10.3390/universe6020021},
   number={2},
   journal={Universe},
   publisher={MDPI AG},
   author={Ashtekar, Abhay},
   year={2020},
   month=jan, pages={21} 
   }

@article{senovilla2012stabilityoperatormotscore,
    author = "Senovilla, Jos\'e M. M.",
    editor = "Bi\v{c}\'ak, Ji\v{r}\'\i{} and Ledvinka, Tom\'a\v{s}",
    title = "{On the Stability Operator for {MOTS} and the 'Core' of Black Holes}",
    eprint = "1210.3731",
    archivePrefix = "arXiv",
    primaryClass = "gr-qc",
    doi = "10.1007/978-3-319-06761-2_27",
    journal = "Springer Proc. Phys.",
    volume = "157",
    pages = "215--222",
    year = "2014"
}

@article{Hawking:1971vc,
    author = "Hawking, Stephen W.",
    title = "{Black holes in general relativity}",
    doi = "10.1007/BF01877517",
    journal = "Commun. Math. Phys.",
    volume = "25",
    pages = "152--166",
    year = "1972"
}

@misc{kehle2024extremalblackholeformation,
      title={Extremal black hole formation as a critical phenomenon}, 
      author={Christoph Kehle and Ryan Unger},
      year={2024},
      eprint={2402.10190},
      archivePrefix={arXiv},
      primaryClass={gr-qc},
      url={https://arxiv.org/abs/2402.10190}, 
}

@article{Booth:2021sow,
    author = "Booth, Ivan and Hennigar, Robie A. and Pook-Kolb, Daniel",
    title = "{Ultimate fate of apparent horizons during a binary black hole merger. I. Locating and understanding axisymmetric marginally outer trapped surfaces}",
    eprint = "2104.11343",
    archivePrefix = "arXiv",
    primaryClass = "gr-qc",
    doi = "10.1103/PhysRevD.104.084083",
    journal = "Phys. Rev. D",
    volume = "104",
    number = "8",
    pages = "084083",
    year = "2021"
}

@article{Andersson:2005gq,
      author         = "Andersson, Lars and Mars, Marc and Simon, Walter",
      title          = "{Local existence of dynamical and trapping horizons}",
      journal        = "Phys.Rev.Lett.",
      volume         = "95",
      pages          = "111102",
      doi            = "10.1103/PhysRevLett.95.111102",
      year           = "2005",
      eprint         = "gr-qc/0506013",
      archivePrefix  = "arXiv",
      primaryClass   = "gr-qc",
      SLACcitation   = "%%CITATION = GR-QC/0506013;%%",
}

@article{pc,
    author = "Chru\'sciel, P. T.",
    title = "{On the global structure of Robinson-Trautman space-times}",
    doi = "10.1098/rspa.1992.0019",
    journal = "Proc. Roy. Soc. Lond. A",
    volume = "436",
    pages = "299--316",
    year = "1992"
}

@article{Podolsky:2009an,
    author = "Podolsky, Jiri and Svitek, Otakar",
    title = "{Past horizons in Robinson-Trautman spacetimes with a cosmological constant}",
    eprint = "0911.5317",
    archivePrefix = "arXiv",
    primaryClass = "gr-qc",
    doi = "10.1103/PhysRevD.80.124042",
    journal = "Phys. Rev. D",
    volume = "80",
    pages = "124042",
    year = "2009"
}

@book{afshordi2024blackholesinside2024,
      title={Black Holes Inside and Out 2024: visions for the future of black hole physics}, 
      eprint={2410.14414},
      pages= {173-197},
      archivePrefix={arXiv},
      primaryClass={gr-qc},
      editor={Afshordi, Niayesh and others},
      Publisher = {arXiv},
      year={2024},
      url={https://arxiv.org/abs/2410.14414} 
}

@incollection{ashtekar2023regularblackholesloop,
      author={Abhay Ashtekar and Javier Olmedo and Parampreet Singh},
      eprint={2301.01309},
      archivePrefix  = {arXiv},
      PrimaryClass = {gr-qc},
      title={Regular black holes from Loop Quantum Gravity}, 
      editor = {Bambi, C.},
      publisher = {Springer},
      booktitle = {Regular Black Holes: Towards a New Paradigm of Gravitational Collapse},
      url={https://arxiv.org/abs/2301.01309}, 
      year={2023}
}

@article{Chen_2022,
   title={Multipole moments on the common horizon in a binary-black-hole simulation},
   volume={106},
   ISSN={2470-0029},
   url={http://dx.doi.org/10.1103/PhysRevD.106.124045},
   DOI={10.1103/physrevd.106.124045},
   number={12},
   journal={Physical Review D},
   publisher={American Physical Society (APS)},
   author={Chen, Yitian and Kumar, Prayush and Khera, Neev and Deppe, Nils and Dhani, Arnab and Boyle, Michael and Giesler, Matthew and Kidder, Lawrence E. and Pfeiffer, Harald P. and Scheel, Mark A. and Teukolsky, Saul A.},
   year={2022},
   month=dec }

@article{Bengtsson_2011,
   title={Region with trapped surfaces in spherical symmetry, its core, and their boundaries},
   volume={83},
   ISSN={1550-2368},
   url={http://dx.doi.org/10.1103/PhysRevD.83.044012},
   DOI={10.1103/physrevd.83.044012},
   number={4},
   journal={Physical Review D},
   publisher={American Physical Society (APS)},
   author={Bengtsson, Ingemar and Senovilla, Jos\'e M. M.},
   year={2011},
   month=feb 
   }

\end{document}